\documentclass{article}
\usepackage{spconf,amsmath,epsfig,amsfonts,amssymb,epsf,enumerate,mathrsfs,amsthm}
\usepackage{psfrag}

              \def\Esp#1{{\mathrm{E}}\bigcro{#1}}

\newsavebox{\fminibox}
\newlength{\fminilength}

  \def\+{^\dagger}

\def\nequiv{\not\kern-.05em\equiv}
\def\egal{\kern-.5em=\kern-.5em}        
\def\propt{\kern-.2em\propto\kern-.2em} 

\def\argmax{\mathop{\mathrm{arg\,max}}} 
\def\argmin{\mathop{\mathrm{arg\,min}}} 
\def\intdouble{\int\kern-0.3em\int}
\def\inttriple{\int\kern-0.3em\int\kern-0.3em\int}

\def\rond#1{\overset{\kern-0.33em~_\circ}{#1}}
\def\rondit[#1]#2{\overset{\kern#1~_\circ}{#2}}

\def\Yb{{\mathbf Y}}
\def\Sb{{\mathbf S}}
\def\Hb{{\mathbf H}}
\def\Zb{{\mathbf Z}}
\def\Hh{{\widehat H}}
\def\Esp{\mathbb{E}}
\title{ACHIEVABLE OUTAGE RATES WITH IMPROVED DECODING OF BICM MULTIBAND OFDM UNDER CHANNEL ESTIMATION ERRORS}
\name{Sajad Sadough$^{\dagger*}$, Pablo Piantanida$^*$, and Pierre Duhamel$^*$}

\address{$^\dagger$Ecole Nationale Sup\'erieure de Techniques Avanc\'ees, 75015 Paris, France\\
Email: sajad.sadough@ensta.fr\\[1mm]
$^*$ Laboratoire des Signaux et Syst\`emes, CNRS/Sup\'{e}lec, F-91192 Gif-sur-Yvette, France\\ Email:\{piantanida, pierre.duhamel\}@lss.supelec.fr \\[1mm]
}

\begin{document}
\ninept
\maketitle
\begin{abstract}
We consider the decoding of bit interleaved coded modulation (BICM) applied to multiband OFDM for practical scenarios where only a noisy (possibly very bad) estimate of the channel is available at the receiver. First, a decoding metric based on the channel {\it a posteriori} probability density, conditioned on the channel estimate is derived and used for decoding BICM multiband OFDM. Then, we characterize the limits of reliable information rates in terms of the maximal achievable outage rates associated to the proposed metric. We also compare our results with the outage rates of a system using a theoretical decoder. Our results are useful for designing a communication system where a prescribed quality of service (QoS), in terms of achievable target rates with small error probability, must be satisfied even in the presence of imperfect channel estimation.   
Numerical results over both realistic UWB and theoretical Rayleigh fading channels show that the proposed method provides significant gain in terms of BER and outage rates compared to the classical mismatched detector, without introducing any additional complexity. 
\end{abstract}
\section{Introduction}
\label{sec:intro}
Ultra-Wide-Band (UWB) is defined as any wireless transmission scheme that occupies a bandwidth of more than 25 \% of its center frequency or greater than 500 MHz over the 3.1-10.6 GHz frequency band \cite{fcc}. Multiband Orthogonal Frequency division multiplexing (MB-OFDM) \cite{batra_jour} is a spectrally efficient technique proposed for high data rate, short range UWB applications. This approach uses a conventional OFDM system, combined with bit interleaved coded modulation (BICM) and frequency hopping for improved diversity and multiple access. In MB-OFDM, the channel is assumed changing so slowly that it is considered time invariant during the transmission of an entire frame.    
Channel estimation is performed by one known symbol (pilot) transmitted at the beginning of the information frame while the rest of the frame is decoded based on the estimated channel. Due to the limited number of pilots, the estimate of the channel is imperfect and the receiver has only access to this noisy channel estimate. However, the receiver/decoder metric for any maximum-likelihood (ML) based detector, requires knowledge of the exact channel.

A standard sub-optimal technique, known as mismatched ML decoding, consists in replacing the exact channel by its estimate in the receiver metric. Hence, the resulting decoding metric is not adapted to the presence of channel estimation errors (CEE). This practical scenario, with mismatched decoding, leads to the following important questions. Firstly, in presence of CEE with a given amount of training, what are the limits of reliable transmission (the capacity). Secondly, what type of encoder/decoder is necessary to transmit reliable information close to the performance limits. The first problem has been recently addressed in \cite{piantanida}, characterizing the maximal reliable information rate, by using the notion of estimation induced outage capacity. Unfortunately, the theoretical encoder/decoder used to achieve this capacity can not be implemented on practical communication systems. Besides, the mismatched ML decoding has been showed to be largely suboptimal for the considered class of channels. 
Basically, the transmitter and the receiver strive to construct codes for ensuring reliable communication with a quality of service (QoS), no matter what degree of channel estimation accuracy arises during the transmission. The QoS requirements stand for achieving target rates with small error probability even with very bad channel estimates. 
In this paper, we propose a practical decoding metric for the aforementioned mismatch scenario. This metric uses the estimated channel and the {\it a posteriori} pdf characterizing the channel estimation process, which matches well the channel knowledge available at the receiver. Based on the derived metric, we formulate our decoding rule for BICM MB-OFDM. Interestingly, the present metric coincides with that derived for space-time decoding from independent results in \cite{Biglieri_jour}. In order to determine the limits of reliable information rates associated to the proposed metric, we use the complementary results obtained in \cite{merhav94}. This allows us to compare the maximal supported rate associated to our metric with that of the classical mismatched ML and the theoretical decoder. Our results are relevant for communication systems where a prescribed quality of service (given by the outage probability) must be ensured even in the presence of imperfect channel estimation. 
  
The outline of this paper is as follows. In Section \ref{sec:sysmodel} we describe the system model for MB-OFDM transmission over a frequency selective fading channel. Section \ref{sec:pilot} presents the pilot assisted channel estimation: we specify the statistics of the CEE and then calculate the posterior distribution of the perfect channel conditioned on the estimated channel. This posterior distribution is used in section \ref{sec:MLmetric} to derive the ML decoding metric in the presence of imperfect channel state information at the receiver (CSIR). In section \ref{sec:dem}, we use the general modified metric for soft decoding  BICM MB-OFDM systems. In section \ref{sec:outageCap}, we derive the achievable outage rates of a receiver using the proposed metric. 
Section \ref{sec:simul} illustrates via simulations the performance of the proposed receiver in realistic UWB channel environments and section \ref{sec:concl} concludes the paper.

Notational conventions are as follows : upper case bold symbols denote vectors, $\mathbf{I}_N$ represents an ($N \times N$) identity matrix; $(.)^T$ and $(.)^{\mathcal{H}}$ denote vector transpose and Hermitian transpose, respectively.   
\section{TRANSMISSION MODEL}
\label{sec:sysmodel}
\begin{figure}[!t]
\centering
\psfrag{A}{\hspace{-1em}Binary Data}
\psfrag{B}{\hspace{-.5em}{\parbox{6em}{\centering $R=1/2$ Convolutional Encoder}}}
\psfrag{C}{\hspace{-1.2em}{\parbox{6em}{\centering Bit\\ interleaver}}}
\psfrag{D}{\hspace{-1.2em}\parbox{6em}{\centering M-QAM\\ Mapping}}
\psfrag{E}{\hspace{-1.2em}\parbox{6em}{\centering IFFT\\ CP \& GI\\ Addition}}
\psfrag{F}{\hspace{-2em}\parbox{6em}{\centering DAC}}
\psfrag{G}{\hspace{-3em}\parbox{6em}{\centering $\exp\left(j2\pi f_ct\right)$}}
\psfrag{H}{\hspace{-.5em}{TFC: Subband Selection}}
\includegraphics[width=0.45\textwidth,height=4cm]{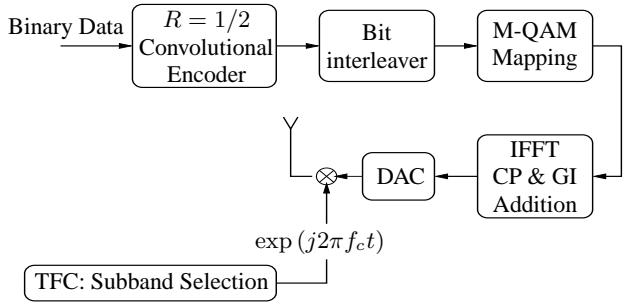}
\caption{TX architecture of the multiband OFDM system.}\label{fig1}
\end{figure}

An MB-OFDM system divides the spectrum between 3.1 to 10.6 GHz into several
non-overlapping subbands each one occupying approximately 500 MHZ of bandwidth
\cite{batra_jour}. Information is transmitted using OFDM modulation over one of the subbands in a particular time-slot. The transmitter architecture is depicted in figure \ref{fig1}: each bit $b_i$ is convolutionally encoded into two bits $c_i^0$ and $c_i^1$ which are interleaved in order to break the error bursts. The interleaved bits are gathered in subsequences of $B$ bits $d_k^1,\ldots,d_k^B$ and mapped to complex M-QAM\footnote[1]{Here, 16-QAM mapping is used instead of QPSK, proposed in \cite{norme_mb}.} ($M=2^B$) symbols $S_k$ with average energy $E_\mathrm{s}=\mathbb{E}[|S_k|^2]$. At a particular time-slot, a time-frequency code (TFC) selects the center frequency of the subband over which the OFDM symbol is sent. The TFC is used not only to provide frequency diversity but also to distinguish between multiple users.  

At the receiver, assuming a cyclic prefix (CP) longer than the channel maximum delay spread and after performing fast Fourier transform (FFT), OFDM converts the the channel into $M$ parallel Rayleigh distributed flat fading subchannels, where $M$ is the number of subcarriers.
The baseband equivalent observation model over a given subband can be written as 
\begin{equation} 
\label{eq:sysmodel}
       \Yb = \mathbf{D}_{_\Hb} \,\Sb + \Zb, 
\end{equation}         
where $(M \times 1)$ vectors $\Yb$ and $\Sb$ denote received and transmitted symbols, respectively; the noise block $\Zb$ is assumed to be a circularly symmetric complex Gaussian random vector with distribution $\Zb \sim \mathcal{CN}(\mathbf{0},{\sigma}^2_z \, \mathbf{I}_M)$; and $\mathbf{D}_{_\Hb}$ is a diagonal matrix with diagonal elements equal to $\Hb=[H_1,\ldots,H_N]^T$, where $\Hb$ is the vector of channel FFT coefficients.  

\section{PILOT ASSISTED CHANNEL ESTIMATION}
\label{sec:pilot}
Under the assumption of time-invariant channel characteristics over the entire transmitted frame, channel estimation is usually performed by transmitting known training symbols at the beginning of each frame. 

Assume that we are interested to estimate the $k$-th, $k=\{ 1, ...,$\\$M\}$, fade coefficient $H_k$ via the transmission of $N$ pilot symbols and let $\Sb_{_T}=[\tilde{S}_1,\ldots,\tilde{S}_N]^T$ be the ($N\times1$) vector of transmitted training sequence. We will assume constant modulus training symbols for the $k$-th subcarrier, i.e., $|\tilde{S}_1|^2 = \ldots=|\tilde{S}_N|^2=P_{T,k}$. From (\ref{eq:sysmodel}) we have 
\begin{equation} 
\label{eq:pil1}
             \Yb_{_T} = \mathbf{D}_{_{\Hb_k}} \, \Sb_{_T} + \Zb_{_T}, 
\end{equation}         
where $\mathbf{D}_{_{\Hb_k}}$ is a diagonal matrix with diagonal elements equal to the $(N \times 1)$ vector \, $\Hb_k = [H_k, \ldots, H_k]^T$ and $\Zb_{_T}$ is a noise vector with the same distribution as $\Zb$ affecting the transmission of pilot symbols. The ML estimate of $H_k$ is obtained by maximizing the likelihood function $p\,(\Yb_{_T}|\Hb_k,\Sb_{_T})$. After some standard calculus we have
\begin{equation}
\label{eq:pil2}
            {\Hh}^{\mathrm{ML}}_k = \frac{\Sb_{_T}^\mathcal{H} \, \Yb_{_T}}{N \, P_{T,k} }=H_k + \mathcal{E}_k, \quad \quad k=1,\ldots,M
\end{equation}
where $\mathcal{E}_k\sim \mathcal{CN}(0,{\sigma}^2_{\mathcal{E}_k})$ is the estimation error. 
From equations \eqref{eq:pil2}, it is obvious that $\sigma^2_{\mathcal{E}_k}=\mathrm{SNR}_{T,k}^{-1}$, where $\mathrm{SNR}_{T,k}\triangleq \frac{N P_{T,k}}{\sigma^2_z}$.

Assuming that $H_k$s are i.i.d. and distributed as $\mathcal{CN}(0,\sigma^2_h)$, the posterior distribution of the perfect channel conditioned on the estimated channel coefficient is given by $p(H_k|\Hh_k)\propto p(\Hh_k|H_k) p(H_k)$.  
After standard manipulation of Gaussian densities we obtain 
\begin{equation}
\label{eq:pil3}
      p(H_k|\Hh_k) = \, \mathcal{CN}\left(\,\rho \Hh_k \,,\frac{\rho \,{\sigma}^2_z}{N P_{T,k}}\right),
\end{equation}
where $\rho_k \triangleq \sigma^2_h /({\sigma}^2_h + \sigma^2_{\mathcal{E}_k} )$.

The availability of the estimation error distribution constitutes an interesting feature of pilot assisted channel estimation that we used to derive the posterior distribution (\ref{eq:pil3}). Next, we see how this additional information is exploited in a modified ML metric for improving the detection process.  
\section{Maximum-likelihood detection metric in the presence of channel estimation errors}
\label{sec:MLmetric}
\subsection{Mismatched ML metric}
The observation model (\ref{eq:sysmodel}) can be rewritten in a component-wise form as
\begin{equation}
\label{eq:ML1}
            Y_k = H_k S_k + Z_k, \quad \quad k=1,\ldots,M
\end{equation}
It is well known that the ML estimate of $S_k$ under the assumption of Gaussian noise is given by maximizing $p(Y_k|H_k,S_k)$ which is equivalent to minimizing the euclidean distance
\begin{equation}
\label{eq:ML2}
            {\widehat{S}}^{\mathrm{ML}}_k(H_k) =\argmin_{S_k \in \, {\mathscr{S}}}\, \big \{\,|Y_k - H_k S_k|^2 \, \big \},
\end{equation} 
where the set $\mathscr{S}$ contains all of the possible discrete values in the constellation that $S_k$ can take.   
Note that the above metric depends on the realization of the {\it perfect} channel $H_k$ and is optimum when perfect CSIR is available. However, in a real communication system, the receiver has only access to an {\it imperfect} estimated version of the channel. In the mismatched decoder, the estimated channel $\widehat{H}_k$ is used instead of $H_k$ in the ML metric of (\ref{eq:ML2}) as 
\begin{equation}
\label{eq:ML3}
            {\widehat{S}}^{\mathrm{ML}}_k(\Hh_k) =\argmin_{S_k \in \, {\mathscr{S}}}\, \big \{\,|Y_k - H_k S_k|^2 \, \big \}_{{\big |_{H_k=\Hh_k}}},
\end{equation} 
which leads to a sub-optimal solution due to the mismatch introduced by the CEE. Next, we will see how the derived posterior channel distribution (\ref{eq:pil3}), can be used in a modified ML metric that takes into account the available imperfect CSI.    
\subsection{Modified ML metric for imperfect CSIR}    
Since in practice the receiver knows solely the imperfect channel, it is relevant to express the likelihood criterion in terms of the estimated fade coefficient. For this purpose, we consider the problem of detecting the transmitted symbol $S_k$ by using the imperfect channel estimate $\widehat{H}_k=\Hh_{k,0}$ in an ML criterion as
\begin{equation}
\label{eq:ML4}
            {\widehat{S}}^{\mathcal{M}}_k(\Hh_{k,0}) =\argmin_{S_k \in \, {\mathscr{S}}} \, \mathcal{D}_{_\mathcal{M}}(Y_k,S_k|\Hh_k=\Hh_{k,0}),
\end{equation} 
where $\mathcal{D}_{_\mathcal{M}}(Y_k,S_k|\Hh_k) \triangleq -\ln p(Y_k|\Hh_k,S_k)$. 
The pdf $p(Y_k|\Hh_k,S_k)$ can be calculated as
\begin{equation}
\label{eq:ML5}
            p(Y_k|\Hh_k,S_k) = \int_{H_k} p(Y_k,H_k|\Hh_k,S_k) \;\; \mathrm{d}H_k,
\end{equation}    
The joint likelihood function ${p(Y_k,H_k|\Hh_k,S_k)}$ is related to known probabilities as
\begin{align}
\label{eq:ML6}
p(Y_k,H_k|\Hh_k,S_k)=p(Y_k|H_k,S_k)p(H_k|\Hh_k),
\end{align} 
where the last equation results from the independence between $S_k$ and $(H_k,\Hh_k)$. 
It is clear from (\ref{eq:ML5}) and (\ref{eq:ML6}) that $p(Y_k|\Hh_k,S_k)$ is obtained by the following conditional expectation 
\begin{equation}
\label{eq:ML7}
            p(Y_k|\Hh_k,S_k) = \mathbb{E}_{H_k |\Hh_k}\big[\,p(Y_k|H_k,S_k)\big |\Hh_k\,\big],
\end{equation}
evaluated over the posterior distribution of the exact channel given its estimate (equation (\ref{eq:pil3})).    
Since $p(H_k|\Hh_k)$  and $p(Y_k|H_k,S_k)$ are Gaussian densities, their product remains Gaussian and it is easy to verify that $p(Y_k|\Hh_k,S_k)=\, \mathcal{CN}\big(\mu_{_\mathcal{M}}\,,{\sigma}^2_{_\mathcal{M}}\big)$ where $\mu_{_\mathcal{M}} =  \rho \, \Hh_k \,S_k$ and $\sigma^2_{_\mathcal{M}} = \sigma^2_z + (1-\rho) \;|S_k|^2$. 
         
According to (\ref{eq:ML4}), the decision metric to be minimized for ML detection at the receiver is easily seen to be
\begin{equation}
\label{eq:ML9}
    \mathcal{D}_{_\mathcal{M}}(S_k,Y_k|\Hh_k)= \ln \,({\sigma^2_z}+(1-\rho)|S_k|^2)+ \frac{|Y_k - \rho \Hh_k S_k|^2}{ {\sigma^2_z}+(1-\rho)|S_k|^2}. 
\end{equation}

Let us first consider the ideal situation when exact channel knowledge is available. In this case $p(H_k|\Hh_k)$ reduces to $\delta(H_k-\Hh_k)$ and the averaging $\mathbb{E}_{_{H_k}}(.)$ of (\ref{eq:ML7}) becomes equivalent to replacing $H_k$ by $\Hh_k$ in $p(Y_k|H_k,S_k)$.  
Similarly, we note that with near perfect channel (i.e. $\sigma^2_\mathcal{E}\to 0$ or $N \to \infty$), $\rho \to 1$ and we have $    \mathcal{D}_{_\mathcal{M}}(Y_k,S_k|\Hh_k) \to |Y_k-\Hh_k S_k|^2, $ which is the euclidean distance metric for perfect CSIR. However, in the presence of CEE, the modified metric evaluates the average of the pdf $p(Y_k|H_k,S_k)$ (that would be used if the exact channel were known) over the true channel $H_k$ using the conditional expectation in (\ref{eq:ML7}). 

While we have focused on OFDM, the metric \eqref{eq:ML9} can also be applied for ML decoding/detection of single carrier systems with frequency selective fading. In the following, we use the above modified metric in the decoding process of BICM MB-OFDM.        
\section{Decoding of BICM-OFDM based on imperfect channel estimation}
\label{sec:dem}
\begin{figure}[!htb]
\centering
\includegraphics[width=0.5\textwidth,height=2.5cm]{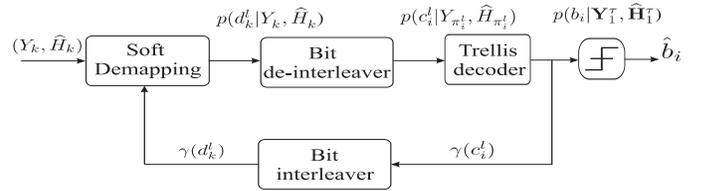}
\caption{BICM iterative demodulation with CEE.}\label{fig2}
\end{figure}

The problem of decoding BICM-OFDM with large size constellations has been addressed in \cite{muq1} under the assumption of perfect CSIR. 
Here we consider the same problem in the case where the decoder has an imperfect estimate of the channel state. 
We denote by $\Sb_1^\tau=[\Sb_1^T,\ldots,\Sb_\tau^T]$ the sequence of transmitted complex symbols and by $\Yb_1^\tau=[\Yb_1^T,\ldots,\Yb_\tau^T]$ the corresponding sequence of received symbols where $\tau$ is number of OFDM symbols in a frame.   
Based on the transmission model of figure \ref{fig1}, the sequence of transmitted bits $\mathbf{b}$ has to be estimated in the ML sense so as to maximize the likelihood function $p(\Yb_1^\tau|\mathbf{b})$. In practice, due to the one to one correspondence between the information bits and the interleaved codeword $\mathbf{d}$, one look for the ML estimate $\widehat{\mathbf{d}}$ of the interleaved codeword as
\begin{equation}
\label{eq:dem1}
        \widehat{\mathbf{d}}^{\mathcal{M}}= \argmax_{\mathbf{d} \in \, {\mathscr{D}}} p(\Yb_1^\tau \big| \widehat{\Hb}_1^\tau,\Sb_1^\tau).
\end{equation}    
By assuming that the received symbols are independent and the channel noise is white, (\ref{eq:dem1}) is equivalent to (see \cite{muq1})
\begin{equation}
\label{eq:dem3}
\widehat{\mathbf{d}}^{\mathcal{M}} \approx \argmax_{\mathbf{d} \in \, {\mathscr{D}}} \prod_k \prod_{l=1}^B p(d_k^l|\Hh_k,Y_k),
\end{equation}
where the bit metrics $p(d_k^l|\Hh_k,Y_k)$ are given by
\begin{align}
     p(d_k^l|\Hh_k,Y_k) & \propto \sum_{\begin{subarray}{l} d_k^n \in \{0,1\} \\ \forall l \neq n \end{subarray}} p(Y_k|\Hh_k,S_k(d_k^1,\ldots,d_k^B)), \notag \\
    & \propto \sum_{\begin{subarray}{l} d_k^n \in \{0,1\} \\ \forall l \neq n \end{subarray}} \frac{1}{ \pi \sigma^2_{_\mathcal{M}} } \, \exp \big (-\frac{|Y_k-\mu_{_\mathcal{M}}|^2}{\sigma^2_{_\mathcal{M}}} \big),
\end{align}
and $p(Y_k|\Hh_k,S_k)$ is specified from \eqref{eq:ML7}.
We emphasize that the above decoding rule is conditioned on the imperfect estimated channel sequence $\widehat{\Hb}_1^\tau$ available at the receiver. Information bits are estimated in the trellis decoder part from the above bit metrics by applying either non-iterative or turbo procedures (see figure \ref{fig2}).    

\section{Achievable Outage Rates Associated to the Proposed Metric}
\label{sec:outageCap}
We now derive the achievable rates $C_{_\mathcal{M}}$ of a receiver using the modified metric $\mathcal{D}_{_\mathcal{M}}$ of \eqref{eq:ML9}. This is done by using the following theorem proposed in \cite{merhav94} for a discrete-time memoryless channel $W(\mathbf{y}|\mathbf{s},\mathbf{H})=\prod_{k=1}^M \mathcal{CN}(H_k \, s_k,\sigma^2_z)$, where $\mathbf{y}=(y_1,\dots,y_M)$,  $\mathbf{s}=(s_1,\dots,s_M)$ with decoding metrics $\mathcal{D}_{_\mathcal{M}}(y_k,s_k | \widehat{H}_k)$: 
\newtheorem{theo}{Theorem}
\begin{theo}
\label{th1}
\begin{equation}
\label{eq:outage}
C_{_\mathcal{M}}=\sup_{p(s)} \; \inf_{f(y|s)\in\mathcal{F}} D(f \!\circ\! P_S \|  P_S  P_Y),
\end{equation}   
where the relative entropy functional $D(f \!\circ\! P_S \|  P_S  P_Y)=$
\begin {equation}
\label{eq:I} 
\int \int P_S(\mathbf{s}) f(\mathbf{y}|\mathbf{s}) \log \frac{f(\mathbf{y}|\mathbf{s})} {\int P_{S^\prime}(s^\prime)f(\mathbf{y}|\mathbf{s}^\prime) \textrm{d}\mathbf{s}^\prime}  \textrm{d}\mathbf{y} \textrm{d}\mathbf{s},
\end{equation} 
and $\mathcal{F}$ denotes the set of all possible channels $f(\mathbf{y}|\mathbf{s})$ that satisfy: 
\begin{align}
&\mathrm{c_1}: \; \Esp_\mathbf{s} \big[f(\mathbf{y}|\mathbf{s})\big]=\Esp_\mathbf{s} \big[W(\mathbf{y}|\mathbf{s},\mathbf{H})\big], \;\;\;\;\;\; \textrm{a.s.}\vspace{-2mm} 
\\  \label{eq:ineqconstr}
&\mathrm{c_2}: \; \Esp_\mathbf{s}  \, \bigg[ \Esp_{_{f(\mathbf{y} |\mathbf{s} )}} \big[\mathcal{D}(s_k,y_k|\widehat{H}_k)\big] \bigg ] \leq \Esp_\mathbf{s} \, \bigg[ \Esp_{_{W(\mathbf{y} |\mathbf{s} )}} \big[\mathcal{D}(s_k,y_k|\widehat{H}_k )\big]\bigg],\vspace{-2mm}
\end{align}
for every $k=1,\dots,M$.
\end{theo}

We assume Gaussian i.i.d. inputs $s_k \sim \mathcal{CN}(0,P)$. Furthermore, since we aim to minimize the mutual information in \eqref{eq:outage}, we choose the Gaussian distribution  $f(\mathbf{y}|\mathbf{s},\underline{\mu})=\prod_{k=1}^M \mathcal{CN}(\mu_k s_k,\sigma^2_k)$ with $\underline{\mu}=(\mu_1,\dots,\mu_M)$. It is easy to see from constraint $\mathrm{c_1}$ that the variances of the optimal channel distribution $f(\mathbf{y}|\mathbf{s},\underline{\mu})$ are given by
\begin{equation}
\sigma^2_k = \sigma^2_z + P(|H_k|^2-|\mu_k|^2),\,\,\,\,k=1,\dots,M.  
\end{equation} 
Then, for any given channel $\mathbf{H}$ and its estimate $\widehat{\mathbf{H}}$, the resulting achievable rates for the proposed metric are obtained as 
\begin{equation}
\label{eq:CM}
C_{_\mathcal{M}}(\mathbf{H},\widehat{\mathbf{H}})=\!\!\!\!\! \!\!\!\!\!\inf_{\underline{\mu}: \,\,\mu_k\in \mathcal{V}(H_k,\widehat{H}_k)}\sum\limits_{k=1}^{M} \log_2\left(1+\frac{P\;|\mu_k|^2}{\sigma^2_k}\right),
\end{equation}
where $\mathcal{V}(H_k,\widehat{H}_k)$ are the constraint sets $\mathrm{c_2}$ corresponding to the metric $\mathcal{D}_{\mathcal{M}}$, derived from the inequality constraint \eqref{eq:ineqconstr} as
\begin{equation}
\label{ineqconstr2}
\mathcal{V}(H_k,\widehat{H}_k)=\big\{\mu_k \in \mathbb{C}:|H_k-a_k \widehat{H}_k |^2 \leq |\mu_k-a\widehat{H}_k|^2\big \},
\end{equation}
where $a_k=\rho_k(\, \lambda_k \sigma^2_z -P(1-\rho_k)\,)/(\, \lambda_k \sigma^2_z-P(1-\rho_k)(1-\lambda_k)\,)$; \\
$\lambda_k=\exp \big( \frac{\sigma^2_z}{P(1-\rho_k)}\big) E_1\big(\frac{\sigma^2_z}{P(1-\rho_k)}\big)$ and $E_1(x) \triangleq \int^{+\infty}_{x} \frac{\exp(-u)}{u} \mathrm{d}u$ is defined as the exponential integral. 

We note that \eqref{eq:CM} is an increasing function of $\|\underline{\mu}\|^2$. Consequently, the infimum is obtained for $\mu_k=\mu_{\mathrm{opt},k}$ resulting from minimizing $\|\underline{\mu}\|^2$ constrained by the equality in the constraint set \eqref{ineqconstr2}. Finally, by using Lagrange multipliers we obtain
\begin{equation}
\mu_{\mathrm{opt},k} = \eta_{_{\mathcal{M},k}} \widehat{H}_k, \,\;\;\,k=1,\dots,M
\end{equation}
where 
\begin{equation}
\eta_{_{\mathcal{M},k}} =\left\{ \begin{array}{ll} a_k-\displaystyle{\frac{|H_k-a_k\widehat{H}_k|}{|\widehat{H}_k|}}, & \mathrm{if}\,\, a_k \geq 0\\
 a_k+\displaystyle{\frac{|H_k-a_k\widehat{H}_k|}{|\widehat{H}_k|}}, & \mathrm{if}\,\, a_k < 0. 
\end{array} \right.
\end{equation}
The achievable rates associated to the metric $\mathcal{D}_{_\mathcal{M}}$ are given by
\begin{equation}
C_{_\mathcal{M}}(\mathbf{H},\widehat{\mathbf{H}})=\sum\limits_{k=1}^M\log_2\left(1+\frac{P\; \eta_{_{\mathcal{M},k}}^2|\widehat{H}_k|^2}{\sigma^2_z + P(|H_k|^2-\eta_{_{\mathcal{M},k}}^2|\widehat{H}_k|^2)}\right).
\end{equation}  
Then, the associated outage probability for an outage rate $R \geq 0$ is defined as
\begin{equation}
P_{_{\mathcal{M}}}^{\mathrm{out}}(R,\widehat{\mathbf{H}})=\mathrm{Pr}_{\mathbf{H}| \widehat{\mathbf{H}}} \big(\mathbf{H}\in \Lambda_{_\mathcal{M}}(R,\widehat{\mathbf{H}})|\widehat{\mathbf{H}}\big), \notag
\end{equation}
with $\Lambda_{_\mathcal{M}}(R,\widehat{\mathbf{H}})=\big\{\mathbf{H}:\, C_{_\mathcal{M}}(\mathbf{H},\widehat{\mathbf{H}})<R\big\}$ and the maximal outage rate for an outage probability $\gamma$ is
\begin{equation}
\label{capout}
C_{_\mathcal{M}}^{\mathrm{out}}(\gamma,\widehat{\mathbf{H}})=\sup_R\big\{R \geq 0: P_{_{\mathcal{M}}}^{\mathrm{out}}(R,\widehat{\mathbf{H}})\leq \gamma\big\}. 
\end{equation}
Since this outage rate still depends on the random channel estimates $\widehat{\mathbf{H}}$, we will consider the average rates over all channel estimates
\begin{equation}
\overline{C}_{_\mathcal{M}}^{\; \mathrm{out}}(\gamma)=\mathbb{E}_{\widehat{\mathbf{H}}}\big[C_{_\mathcal{M}}^{\mathrm{out}}(\gamma,\widehat{\mathbf{H}})\big]. \label{EQ-capacity}
\end{equation}
The achievable rates \eqref{EQ-capacity} are upper bounded by the estimation-induced outage capacity, the maximal outage rates (i.e. maximizing over all possible transmitter-receiver pairs), derived in \cite{piantanida} by using a theoretical decoder. In our case, this capacity is given by
\begin{equation}
\overline{C}_{_\mathcal{G}}^{\; \mathrm{out}}(\gamma)=\mathbb{E}_{\widehat{\mathbf{H}}}\big[C_{_\mathcal{G}}^{\mathrm{out}}(\gamma,\widehat{\mathbf{H}})\big], \label{perfect-capacity}
\end{equation}
where the outage rates $C_{_\mathcal{G}}^{\mathrm{out}}(\gamma,\widehat{\mathbf{H}})$ are computed using 
\begin{equation}
C_{_\mathcal{G}}(\mathbf{H},\widehat{\mathbf{H}})=\sum\limits_{k=1}^{M} \log_2\left(1+\frac{P\;|H _k|^2}{\sigma^2_z}\right). 
\end{equation}

\begin{figure}[!htb] 
\centering
\includegraphics[width=0.5\textwidth,height=0.28\textheight]{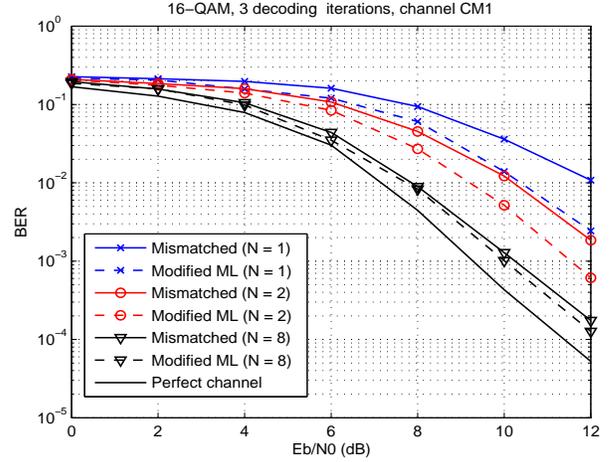} 
\caption{BER performance of the proposed decoder over the LOS UWB channel CM1 for various training sequence lengths.}\label{fig3}
\end{figure} 

\begin{figure}[!htb] 
\centering
\includegraphics[width=0.5\textwidth,height=0.28\textheight]{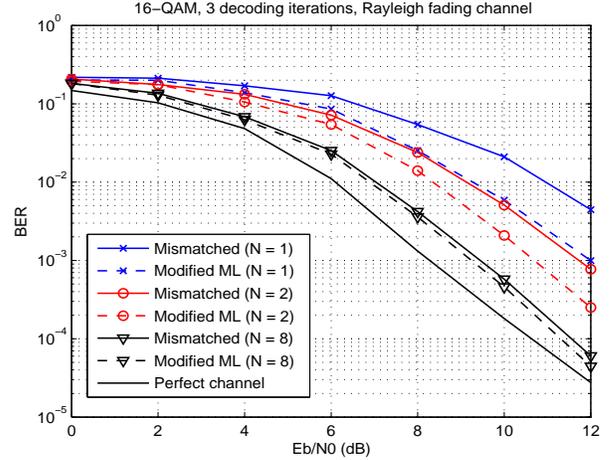} 
\caption{BER performance of the proposed decoder over Rayleigh fading channel for various training sequence lengths.}\label{fig4}
\end{figure} 
\section{Simulation results}
\label{sec:simul}
Simulations have been carried out in the context of IEEE802.15 wireless PAN \cite{norme_mb}: $M=100$ data subcarriers along with 32 CP samples compose one OFDM symbol. 
Information bits are encoded by a rate $R=1/2$ convolutional encoder with constraint length 3 defined in octal form by (5,7). The interleaver is a pseudo-random one operating over the entire frame and the mapping is 16-QAM. 
For each frame that contains 100 OFDM symbols, a different realization of the realistic UWB channel model specified in \cite{chanReport} has been drawn and supposed constant during the whole frame. Besides, the same average energy is allocated to training symbols and data, i.e., $E_{_T}=\mathbb{E}(|\tilde{S}_k|^2)=E_\mathrm{s}$. 

Figure \ref{fig3} depicts the bit error rate (BER) performance gain that is obtained by decoding BICM MB-OFDM with the modified ML decoder in the context of LOS CM1 channel, where the channel estimation is performed by sending different training sequences with lengths $N=\{1, 2, 8\}$ per frame. 
It can be noticed that the proposed decoder outperforms the mismatched decoder especially when few numbers of pilot symbols are dedicated for channel estimation. Note that the modified decoder with $N=1$ pilot performs very close to the mismatched decoder with $N=2$ pilots. 
For comparison, results obtained with theoretical Rayleigh fading channel are illustrated in figure \ref{fig4}. It can be observed that for $N=2$, the SNR to obtain a BER of $10^{-3}$ is reduced by about $1.5$ dB if the modified ML decoding is used instead of the mismatched approach.     

\begin{figure}[!htb] 
\centering
\includegraphics[width=0.5\textwidth,height=0.28\textheight]{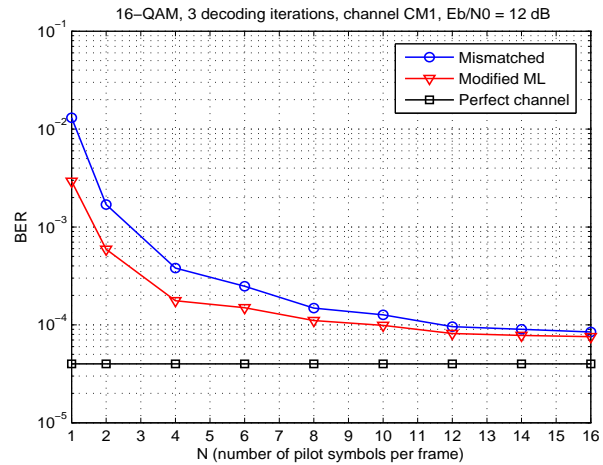} 
\caption{BER versus the number of pilot symbols at $E_b/N_0=12$ dB over the LOS channel CM1.}\label{fig5}
\end{figure}

Figure \ref{fig5} shows the BER performance versus the training sequence length $N$ at a fixed $Eb/N_0$ of $12$ dB for the CM1 channel. This allows to evaluate the number of training sequence necessary to achieve a certain BER. We observe that at $Eb/N_0=12$ dB, the modified ML decoder requires $N=9$ pilot symbols per frame to achieve a BER of $10^{-4}$ while the mismatched decoder attains this BER for $N=12$ pilot symbols. Besides, this has been outlined in Section IV, for large training sequence lengths ($N \geq 12$), both decoders have close performance.   

\begin{figure}[!htb] 
\centering
\includegraphics[width=0.5\textwidth,height=0.27\textheight]{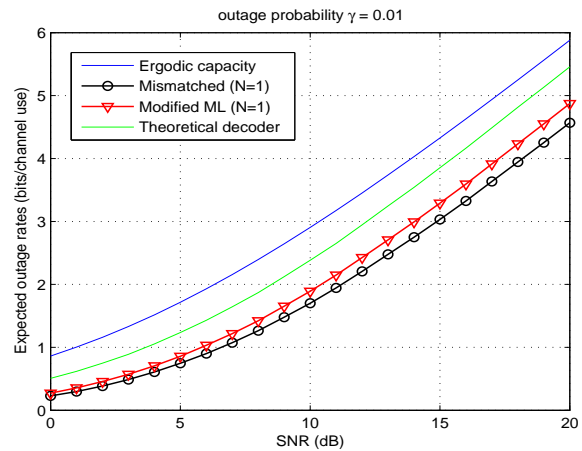} 
\caption{Expected outage rates of OFDM transmission with $M=100$ subcarriers versus SNR $(N=1)$.} \label{fig6}
\end{figure}

Figure \ref{fig6} shows the expected achievable rates \eqref{EQ-capacity} corresponding to one OFDM symbol transmission with $M=16$ data subcarriers (in bits per channel use) versus the SNR, for an outage probability $\gamma=0.01$. For comparison, we show the expected outage rates of mismatched ML decoding and that obtained in the case of the perfect theoretical encoder/decoder \eqref{perfect-capacity}. It can be observed that at a mean outage rate of 4 bits, the modified metric reduces the SNR by about 1 dB.  Indeed, we observe that the achievable rates using the proposed decoding metric are close (only $2$ dB of SNR) to those obtained by using a theoretical decoder.             

\section{Conclusion}
\label{sec:concl}
This paper studied the problem of ML reception when the receiver has only access to a noisy estimate of the channel in the case of pilot assisted channel estimation. 
By using the statistics of the estimation error, we proposed a modified ML criterion that is expressed in terms of the estimated channel coefficient.  
This modified metric let us to formulate appropriate branch metrics for decoding BICM MB-OFDM with imperfect channel knowledge. Achievable outage rates associated to this modified metric have also been derived. Comparison in terms of BER and achievable outage rates was made with a system whose receiver uses mismatched ML decoding and a perfect theoretical decoder. 
Simulation results conducted over realistic UWB channels, indicate that mismatched decoding is quite sub-optimal for short training sequence and confirmed the adequacy of the proposed decoding rule in the presence of channel estimation errors. This was obtained without introducing any additional complexity.         
\bibliographystyle{IEEEbib}
\bibliography{bibli_asilomar}

  
\end{document}